# An iterative method for reference pattern selection in high-resolution electron backscatter diffraction (HR-EBSD)

Abdalrhaman Koko[1a,b], Vivian Tong[b], Angus J. Wilkinson[a], and T. James Marrow[a]

a    Department of Materials, University of Oxford, Oxford OX1 3PH, United Kingdom

b    National Physical Laboratory, Hampton Road, Teddington, TW11 0LW, United Kingdom

## Abstract

For high (angular) resolution electron backscatter diffraction (HR-EBSD), the selection of a reference diffraction pattern (EBSP$_0$) significantly affects the precision of the calculated strain and rotation maps. This effect was demonstrated in plastically deformed body-centred cubic and face-centred cubic ductile metals (ferrite and austenite grains in duplex stainless steel) and brittle single-crystal silicon, which showed that the effect is not only limited to measurement magnitude but also spatial distribution. An empirical relationship was then identified between the cross-correlation parameter and angular error, which was used in an iterative algorithm to identify the optimal reference pattern that maximises the precision of HR-EBSD.

---

[1] Corresponding author. E-mail address: abdo.koko@materials.ox.ac.uk

# 1. Introduction

Electron backscatter diffraction (EBSD) technique offers an impressive combination of sensitivity, spatial resolution and ease of use compared to other methods in measuring the elastic strains and estimating the geometrically necessary dislocation (GND) density which provides quantitative information about a material's elastic and plastic behaviour at the microscale. [1–3]. For electron backscatter diffraction microscopy, a flat polished crystalline specimen is placed inside a scanning electron microscope (SEM) [4]. Tilting the sample elongates the interaction volume perpendicular to the tilt axis, allowing more backscattered electrons to leave the sample [5,6]. The electron beam (typically 20 kV) is focused on a small volume and backscatters at a spatial resolution of ~20 nm at the specimen surface [7]. The spatial resolution varies with angular width [8], interaction volume [9], and nature of the material under study [7] (in the related technique of transmission Kikuchi diffraction, the resolution also varies with the specimen thickness [10]. Thus, increasing the beam energy increases the interaction volume and decreases the spatial resolution [11].

Rastering the beam position to obtain EBSD maps can provide information about texture, grain size, misorientation across boundaries (e.g., grain or twin), and local misorientations of the crystal [1,12,13]. In addition, finer features (e.g., phase, polarity) can be adequately quantified from the intensity distribution within the Kikuchi bands [14]. Furthermore, the change and degradation in electron backscatter patterns (EBSPs) provide information about the diffracting volume. Pattern degradation (i.e., diffuse quality) can be used to assess the level of plasticity through the pattern or image quality (IQ) [15], where IQ is calculated from the sum of the peaks detected when using the conventional Hough transform [16]. Wilkinson [17] first used the changes in high-order Kikuchi line positions to determine elastic strains, albeit with low precision[2] (0.3% to 1%); however, this approach cannot be used for characterising residual elastic strain in metals as the elastic strain at the yield point is usually around 0.2%. Measuring strain by tracking the change in the higher-order Kikuchi lines is practical when the strain is small, as the band position is sensitive to changes in lattice parameters [18]. In the early 1990s, Troost *et al.* [19] and Wilkinson *et al.* [20,21] used pattern degradation and change in the zone axis position to measure the residual elastic deviatoric strains and small lattice rotations with a 0.02% precision.

---

[2] Throughout the paper, the terms 'error', and 'precision' were used as defined in International Bureau of Weights and Measures (BIPM) guide to measurement uncertainty (GUM). The true value in any measurement is in practice unknowable, therefore treat 'error', 'accuracy', and 'uncertainty' as synonymous, and 'true value' and 'best guess' as synonymous. Precision is the variance (or standard deviation) between all estimated quantities. Bias is the difference between average of measured values and an independently measured 'best guess'. Accuracy is then the combination of bias and precision.

Cross-correlation-based, high angular resolution electron backscatter diffraction (HR-EBSD) – introduced by Wilkinson *et al.* [22,23] – is a scanning electron microscopy (SEM) -based technique to map relative deviatoric elastic strains and rotations, and estimate the geometrically necessary dislocation (GND) density in crystalline materials. HR-EBSD method uses image cross-correlation to measure pattern shifts between regions of interest (ROI) between electron backscatter diffraction patterns (EBSPs) with sub-pixel precision. As a result, the relative lattice distortion between two points in a crystal can be calculated using pattern shifts from at least four non-collinear ROI. In practice, pattern shifts are measured in more than 20 ROI per EBSP to find a best-fit solution to the deformation gradient tensor, representing the relative lattice distortion[3] [24,25] [22,23]. However, these measurements do not provide information about the volumetric/hydrostatic strains. Full details of the HR-EBSD method are given in [24,26].

Elastic strain and (elastic) lattice rotation tensors are calculated by decomposing the deformation gradient tensor into symmetric and anti-symmetric parts, respectively. Non-hydrostatic components of the residual stress are determined from the elastic strain tensor using Hooke's law with anisotropic elastic stiffness constants. The hydrostatic elastic strain component can also be estimated by assuming that the stress normal to the surface ($\sigma_{33}$) is zero (i.e., a traction-free surface [27]), [23]. Furthermore, the geometrically necessary dislocation (GND) density can be estimated from the HR-EBSD measured lattice rotations by relating the rotation axis and angle between neighbour map points to the dislocation types and densities in a material using Nye's tensor [28,29].

---

[3] Strain, distortion, and deformation can refer to several quantities in different fields. Therefore, we define our use of these terms (in italics) as follows. A mechanically loaded object changes shape in response to applied load; when measured in a mechanical test frame, it is called (total) engineering strain. Plastic strain is called the shape change that persists after the macroscopic load is removed. On the microscale, plastic deformation in most crystalline materials is accommodated by dislocation glide and deformation twinning. However, dislocations are also generated in a material as plastic deformation progresses, and dislocations with similar crystallographic character and sign that end up near each other in a material (e.g., lined up at a slip band) can be characterised as geometrically necessary dislocations (GNDs). Increasing plastic strain in a polycrystal also elastically distorts the crystal lattice to accommodate crystal defects (e.g., dislocation cores), groups of defects (e.g., dislocation cell walls), and maintains compatibility at polycrystal boundaries. This lattice distortion can be expressed as a deformation gradient tensor, which can be decomposed into elastic strain (symmetric) and lattice rotation (antisymmetric) components [100]. In this work, we use the term lattice distortion as a general term to refer to elastic distortion components derived from the deformation gradient, elastic strain, and lattice rotation tensors.

The HR-EBSD method was shown [23,30–32] to achieve a precision of $\pm 10^{-4}$ in components of the displacement gradient tensors (i.e., strain and rotation in radians) by measuring the shifts at a pattern image resolution of $\pm 0.05$ pixels. Still, it was limited to small strains and rotations (>1.5°). Britton and Wilkinson [24] raised the rotation limit to $\approx 11°$ by using a re-mapping technique [33] that recalculated the strain after transforming the patterns with a rotation matrix calculated from the $1^{st}$ cross-correlation iteration. However, further lattice rotation, typically caused by severe plastic deformation, will cause errors in the elastic strain calculations. Ruggles *et al.* [34] demonstrated an improved HR-EBSD precision, even at 12° of lattice rotation, using the inverse compositional Gauss–Newton-based (ICGN) method instead of cross-correlation. Vermeij and Hoefnagels [35] also established a method that achieves a precision of $\pm 10^{-5}$ in the displacement gradient components using a full-field integrated digital image correlation (IDIC) framework instead of dividing the EBSPs into small ROIs. Patterns in IDIC are distortion-corrected to negate the need for re-mapping up to $\approx 14°$ [36,37].

Nonetheless, in HR-EBSD analysis, the lattice distortion field is still calculated relative to a reference pattern or point (EBSP$_0$) per grain in the map, and is dependent on the lattice distortion at the point. The lattice distortion field in each grain is measured with respect to this point; therefore, the absolute lattice distortion at the reference point (relative to the unstrained crystal) is excluded from the HR-EBSD elastic strain and rotation maps [38–40]. This 'reference pattern problem' is similar to the 'd$_0$ problem' in X-ray diffraction [41,42], and affects the nominal magnitude of HR-EBSD stress fields. However, selecting the reference pattern (EBSP$_0$) plays a key role, as severely deformed EBSP$_0$ adds phantom lattice distortions to the map values, thus, decreasing the measurement precision [38,39].

The use of simulated reference patterns for absolute strain measurement is still an active area of research [14,43–50] and scrutiny [38,50–55] as difficulties arise from a variation of inelastic electron scattering with depth which limits the accuracy of dynamical diffraction simulation models, and imprecise determination of the pattern centre which leads to phantom strain components which cancel out when using experimentally acquired reference patterns. Other methods assumed that absolute strain at $EBSP_0$ can be determined using crystal plasticity finite-element (CPFE) simulations, which then can be then combined with the HR-EBSD data (e.g., using linear 'top-up' method [56,57] or displacement integration [58]) to calculate the absolute lattice distortions.

In addition, GND density estimation is nominally insensitive to (or negligibly dependent upon [59,60]) $EBSP_0$ choice, as only neighbour point-to-point differences in the lattice rotation maps are used for GND density calculation [61,62]. However, this assumes that the absolute lattice distortion of $EBSP_0$ only changes the relative lattice rotation map components by a constant value which vanishes during derivative operations, i.e., lattice distortion distribution is insensitive to $EBSP_0$ choice [40].

Existing criteria for $EBSP_0$ selection include: (1) points with low GND density or low Kernel average misorientation (KAM) [63] based on the Hough measured local grain misorientations; (2) points with high image quality (IQ), which may have a low defect density within its electron interaction volume, and is therefore assumed to be a low-strained region of a polycrystalline material [64]. However, IQ does not carry a clear physical meaning [65], and the magnitudes of the measured relative lattice distortion are insensitive to the IQ of $EBSP_0$ [40]; (3) $EBSP_0$ can also be manually selected to be far from potential stress concentrations such as grain boundaries, inclusions, or cracks [40] using subjective criteria. These criteria assume these parameters can indicate the strain conditions at the reference point, producing accurate measurements of up to $3.2 \times 10^{-4}$ elastic strain [31]. However, experimental measurements point to the inaccuracy of HR-EBSD in determining the out-of-plane shear strain components distribution and magnitude [66].

Here, a thorough investigation of the effect of the $EBSP_0$ on the elastic strain, rotation and GND density maps – including magnitude and distribution – was conducted for both brittle and ductile crystal systems. This work introduces a new objective method to select an $EBSP0$ that maximises HR-EBSD measurement precision. The main difference is that the new method considers HR-EBSD quality metrics for all points that use the $EBSP_0$ as a reference. In contrast, existing methods consider conventional EBSD quality metrics related to the $EBSP_0$ alone.

## 2. Methodology.

### 2.1. Materials and EBSD mapping

Two different materials, duplex stainless steel and silicon, were used to provide samples representing a face centre cubic ductile crystal (austenite, high plastic strain), a body centre cubic ductile crystal (ferrite, low plastic strain) and a cubic single-crystal with very low dislocation density, below the noise floor of HR-EBSD lattice rotation measurements (semiconductor-grade silicon). All are model materials for EBSD analysis as the polished surface does not oxidise further, and high-quality patterns can be acquired in a reasonable time.

The 1st set of samples was taken from a large (~200 mm thickness) forging of Zeron 100 duplex stainless steel (UNS: S32760 [67]) with a nominal composition of 25% Cr, 7% Ni, 3.6% Mo, 0.7% Cu, 0.7% W, 0.22% N that was aged at 475 °C for 100 hrs in air. This duplex stainless steel contains face-centred cubic austenite in a matrix of a body-centred cubic ferrite with a volume fraction of 58% [68]. During this heat treatment, the decomposition of the ferrite into Fe-rich nanophase ($\alpha'$) and Cr-rich nanophase ($\alpha''$) occurs with G-phase precipitation [69–71], which increases the ferrite hardness while the hardness of the austenite phase remains unchanged [72–74]. This gives a sample with both face-centre cubic (austenite) and body-centre cubic (ferrite) phases [75].

EBSP quality is extremely sensitive to surface preparation [76]. The aged stainless steel specimen surfaces were prepared by grinding on SiC papers (240 to 4000 grit), polishing using diamond paste (from 9 to 1 μm) and then 50 nm colloidal silica (2 hours, 50 r.p.m speed and 5 N force) to produce a mirror-flat surface without artefacts. The specimens were ultrasonically cleaned for 20 minutes using ethanol, rinsed with deionised water, and dried with a hot air blower.

The single-crystal silicon samples did not require surface preparation. Instead, they were cleaved from a pre-polished (001) single-crystal silicon wafer with a thickness of 0.5 mm. Silicon was selected as it deforms elastically with no plasticity at room temperature [77].

The samples of both materials were positioned between the jaws of a 2 kN 70° pre-tilted loading stage (Deben® MT2000E) inside a Carl Zeiss Merlin field emission gun scanning electron microscope (FEG-SEM). The SEM chamber and loading stage were plasma cleaned and purged together before the sample was deformed in displacement control by the movement of one jaw. The Si samples were compressed to grow a brittle crack, and the stainless steel samples were plastically deformed using uniaxial tension and three-point bending.

EBSD maps were collected from the in-situ strained sample using a 1600 × 1200 pixel Bruker eFlash CCD camera with no additional lens distortion correction. Before each observation, the setup was left to stabilise for 30 minutes at fixed crosshead displacement. The microscope conditions were 10 nA/20 kV beam current/voltage and 18 mm working distance for all materials. EBSPs were saved as an 800 × 600 pixel image, 16-bit depth, with 100 millisecond exposure time per pattern, and a step size of 75 nm for the stainless steel and 250 nm for the silicon. These conditions provided a practical time for measurement while minimising sample drift [28,59,78–80].

## 2.2. EBSP$_0$ effect analysis

After excluding points with a geometrical necessary dislocation (GND) density higher than the grain average (after removing the outliers), which was calculated using MTEX [29]; the reference EBSP$_0$ for each grain can be selected as a point that is remote to stress concentrations as indicated by the GND density (or KAM instead) map that has a high-quality pattern, low GND density, and an orientation close to the grain's mean orientation. EBSP$_0$ selected using this method will be termed '*Native*' in this work. Then, 50 points were randomly selected from a pool of ~15% of the highest-quality patterns to evaluate the effect of EBSP$_0$ selection on the measured HR-EBSD data. The IQ of these randomly selected patterns varied by less than 2% of their average value. Each of the 50 patterns was used in HR-EBSD calculations to map the grain's elastic strain, rotation and GND density distributions.

HR-EBSD analysis was performed using in-house MATLAB software (XEBSD) [24]. EBSP re-mapping was used to minimise strain errors due to large misorientations (>1.5°); in the second pass, the EBSPs were re-mapped to an orientation close to EBSP$_0$ by using the local rotation matrix estimated from the first pass [81]. The elastic constants (in GPa) for the ferrite ($\alpha$) are: $C_{11} = 230, C_{44} = 117, C_{12} = 135$, for the austenite ($\gamma$) are: $C_{11} = 231.4, C_{44} = 116.4, C_{12} = 134.7$ [82], and for (001) silicon are $C_{11} = 165.7, C_{44} = 79.6, C_{12} = 63.9$ [83]; all were transformed to the crystal frame of reference via the Euler angles [84]. The pattern centre (PC) shift due to beam movement during acquisition was corrected using AstroEBSD [85], and 30 ROI were selected from each EBSP for cross-correlation. The ROI size is 256 x 256 pixels, and the calibrated EBSD pixel size is 18 μm. No further angular effect from drift was expected as the field of view was small (~20 x 15 μm$^2$), and the samples were all conductive [86]. GND density was estimated from the local lattice curvature using the method implemented by Wilkinson and Randman [28].

# 3. Results

## 3.1. Body Centre Cubic Ferrite

The aged duplex stainless steel (DSS) sample was deformed in uniaxial tension, and then a ferrite grain was EBSD characterised to investigate the effect of the reference pattern. Figure 1a shows the grain's elastic field (in-plane shear strain) calculated relative to six reference patterns. These six references all have a high IQ compared to other available patterns in the grain. Four points were selected randomly from the subset of high IQ; the fifth is the '*Native*' reference point, as discussed above. The sixth is the optimal point '*Chosen*' by the iterative selection method developed in this work, and the logic behind this will be explained subsequently. The locations of the '*Native*' and '*Chosen'* points are shown in Figure 2.

Using different $EBSP_0$ as a reference influenced the resultant in-plane shear strain distribution. The $1^{st}$ selected point (Figure 1a, point I) is near a low-angle grain boundary, where the grain misorientation is < 0.5°. Despite this low angle, the region near the subgrain boundary, where the $EBSP_0$ has been selected from, is likely to have high localised strains (in this case, a positive in-plane shear strain); thus, its selection as a reference causes the relative strains measured in the grain to be shifted towards the negative direction. This is illustrated by the line profile (dark blue in Figure 1b), which shows that the in-plane shear strain magnitude is shifted towards negative shear strains compared to the other references. The $2^{nd}$ point (Figure 1a, point II) is at the centre of the grain and leads to significantly more positive strains (light blue line profile in Figure 1b). The $3^{rd}$ point is close to a linear feature (i.e., deformation twin) in the grain (marked by the arrow in Figure 1b), and the $4^{th}$ point (Figure 1a.III) has the highest IQ in the grain. These references give strain profiles intermediate between points I and VI, as does the $5^{th}$ reference point (V), the '*Native'* reference. The $6^{th}$ '*Chosen*' reference gives similar strains to point III. Generally, the difference between the lines profile is not due to an offset but noticeable non-linear changes in the distribution.

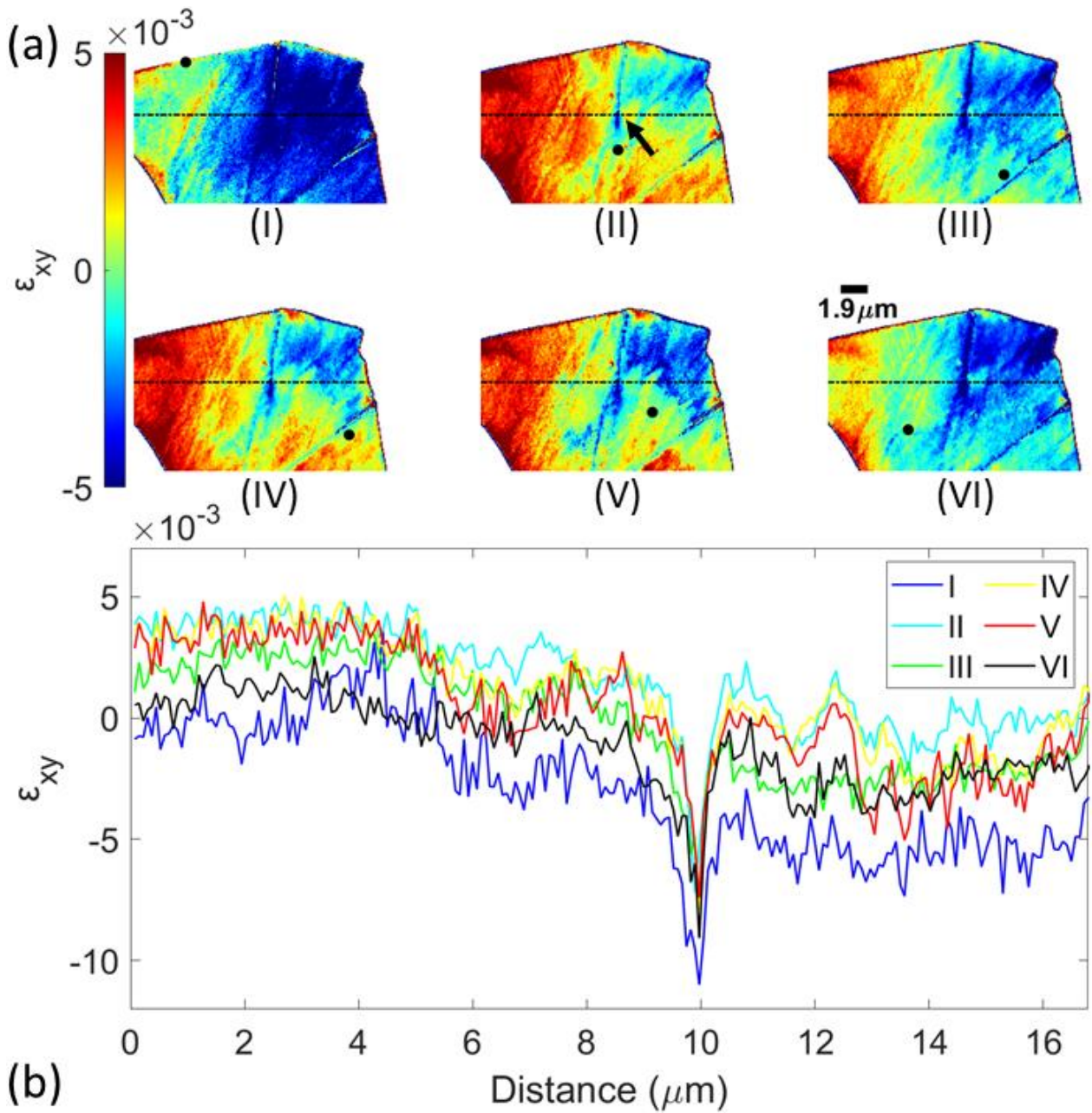

Figure 1: a) In-plane shear strain ($\varepsilon_{xy}$) fields in a ferrite grain, obtained using six different patterns as the reference EBSP$_0$. A black dot indicates the position of the reference pattern in each case. b) The line profile of the in-plane shear strain along the dotted line. The black arrow points to the value ahead of the linear feature (i.e., deformation twin) located in the middle of the grain (see Figure 2). V and VI are the '*Native*' and '*Chosen*' EBSP$_0$, respectively. The IQ values of points I to VI are 0.4019, 0.4419, 0.4288, 0.4563, 0.4400 and 0.4419. The sample is at 15% engineering strain, as detailed in reference [87].

The set of 50 randomly selected high-quality reference points with a mean of 0.4384 ± 0.0090, in addition to the '*Native*' point with an IQ of 0.4400 and the '*Chosen*' point with an IQ of 0.4419, marked in Figure 2, which also shows the GND density map superimposed on the forescatter electron image. The high GND density along the deformation twin boundaries is apparent. The points of high IQ are generally at locations with low GND density (see the supplementary information: A). These references were used individually to produce 52 HR-EBSD maps in total.

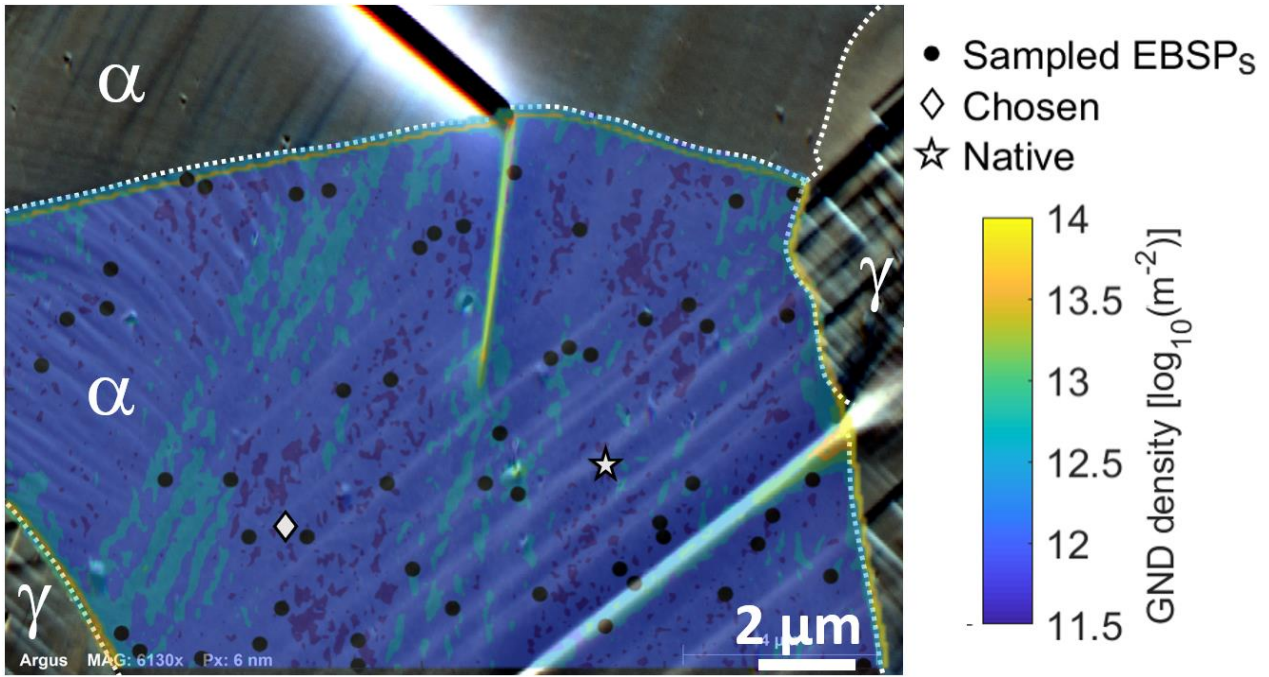

Figure 2: Geometrical necessary dislocation (GND) density map overlaid over a forescatter electron (FSD) image for the field of view with the location of 52 EBSPs (50 random + *Native* + *Chosen*) in the example ferrite grain. One '*Native'*-ly selected (white star), 50 were randomly '*Sampled*' (black points), and another '*Chosen*' (white diamond). The average GND density is 11.85 ± 0.74 $\log_{10}(m^{-2})$.

The effect of the local conditions at each $EBSP_0$ can be observed by considering the 52 HR-EBSD maps obtained from the 50 high-quality reference patterns randomly selected plus '*Native'* and '*Chosen'* EBSPs. To do this, the average strains in the HR-EBSD map arising from each $EBSP_0$ were compared with the conditions at the $EBSP_0$ in the 51 other HR-EBSD maps. These results are shown in Figure 3. Figure 3 shows the relationship between the grain's average strains relative to a single $EBSP_0$ ($\varepsilon^{Grain}$, horizontal axis), and the average strain measured at this $EBSP_0$ point when using the other 50 $EBSP_0$ ($\varepsilon^{EBSP0}$, vertical axis). All three strain components show an inverse linear correlation; this is expected because the strains at the grain are measured relative to the strain state at $EBSP_0$, which clearly shows the HR-EBSD map's dependency on the reference pattern choice, even from a set of high-quality references. The '*Chosen*' $EBSP_0$ is near the middle of the strain distribution for all three strain components; this most likely contributes to a high cross-correlation precision as it minimises the distortions between the $EBSP_0$ and other points in the grain to around half of the strain distribution range. In contrast, the '*Native*' $EBSP_0$ is near the middle of the strain distribution range for $\varepsilon_{xx}$ and $\varepsilon_{xy}$ components but at the negative end of the

ε_yy strain distribution range, which leads to a larger maximum ε_yy distortion with correspondingly higher measurement uncertainties.

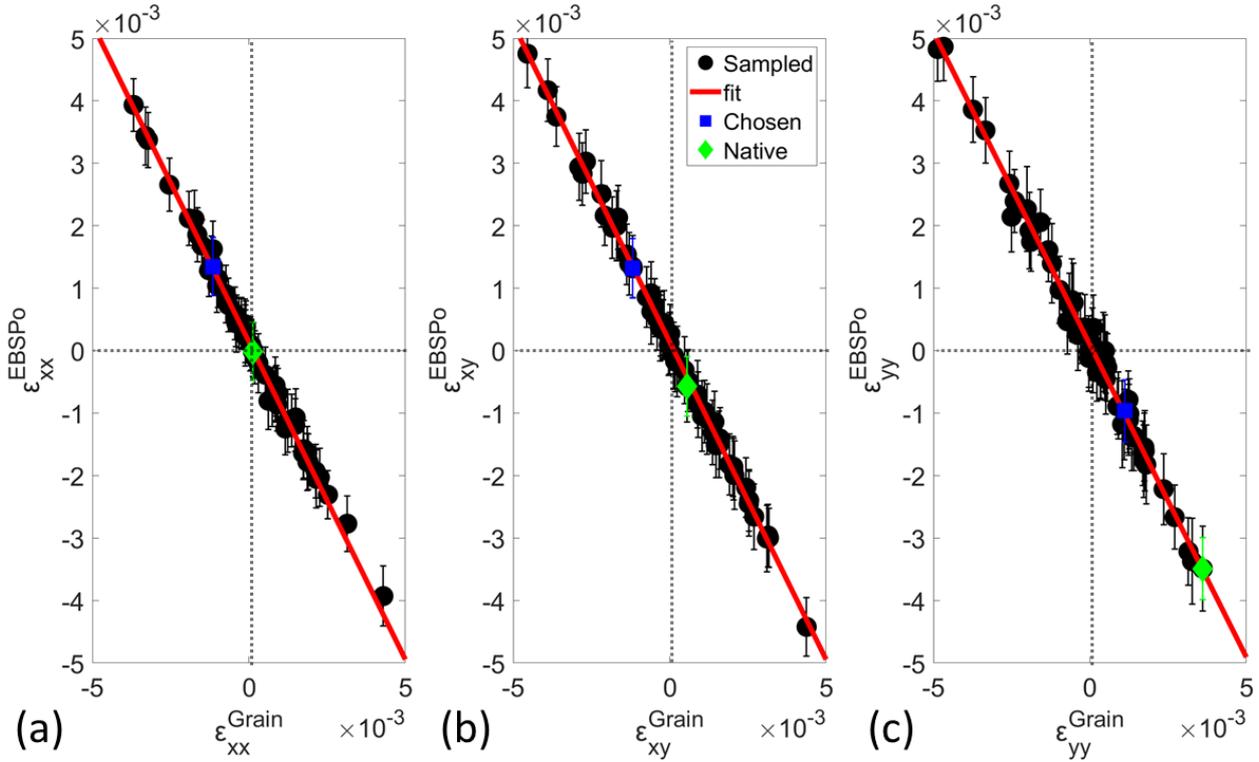

Figure 3: Grain (field value) and EBSP₀ (local value) averaged (a) $\varepsilon_{xx}$, (b) $\varepsilon_{xy}$, and (c) $\varepsilon_{yy}$ strains for the selected reference patterns with a correlation coefficient of 0.99 across the three maps. Only the standard deviation in the EBSP₀ local value is shown as the standard deviation for all the points in the grain is large.

The correlation coefficients between the local conditions at the EBSP₀ point, as averaged across the other HR-EBSD maps calculated using different EBSPs (labelled EBSP₀), and the conditions at the grain, as averaged from all the points within the grain (labelled 'Grain'), are summarised in Figure 4; for parameters that include: the lattice distortion components, e.g., determinant of the deformation gradient tensor ($A^o$), in equation (1) where $n$ is the total of points, strain $\varepsilon_{ij}$, rotation $\omega_{ij}$, and GND density, IQ, PH which is the (mean) cross-correlation peak height normalised by the value from the reference self-correlation in each case, and the mean angular error (MAE), which is a quantitative measure of the difference between the pattern shift at each segmented region (ROI) from the EBSP, and the best (least-square) fit solution, all obtained after re-mapping. In general, the lower the MAE, the higher the precision of the solution [81].

$$A^o = \frac{1}{n}\sum_{el=1}^{n}\left|F_{ij}\right|_{el}, \qquad i = j = x, y, z \qquad 1$$

The correlations were quantified through Pearson's correlation coefficient [88] implemented in the MATLAB® *corrcoef* function (eq. 2). Each dataset $(A, B)$ was first normalised using the mean $(\sigma)$ and standard deviation $(\mu)$ before finding the linear correlation coefficient $(\rho)$ of two datasets with $N$ number of observations.

$$\rho(A, B) = \frac{1}{N-1} \sum_{i=1}^{N} \left( \frac{A_i - \mu_A}{\sigma_A} \right) \left( \frac{B_i - \mu_B}{\sigma_B} \right) \qquad 2$$

The same analysis was applied to EBSD maps in 8 different grains of the body-centred cubic ferrite (386 HR-EBSD maps, at different engineering strains – see the supplementary information: A). The correlation coefficients between the $EBSP_0$ and the grain are presented in Figure 4 by averaging the correlation coefficients from 9 sets of analyses.

The diagonal of Figure 4 shows a strong inverse correlation between the field and local reference pattern status in all of the strain and rotation tensor components because the grain's average lattice distortion components are measured relative to the local $EBSP_0$ values. The correlation is stronger for the rotations than for the strains because measurement noise forms a higher proportion of the measured elastic strains than for lattice rotations. Typical lattice rotations produce EBSP ROI shifts, which are an order of magnitude larger than elastic strains. This indicates that a reference pattern deformed in tension will directly reduce the tensile strain magnitudes of the resultant map while indirectly influencing the other component magnitude and the strain's distribution.

The lattice distortion components (in 'Grains' and '$EBSP_0$') are insensitive to the local cross-correlation quality metrics (i.e., PH and MAE, magenta box annotation in Figure 4 for 'Grains' components). However, the correlation is limited to linear relationships. Significantly, PH and MAE both have positive diagonal correlation coefficients (yellow boxes in Figure 4), which shows that $EBSP_0$ with high local PH tend to produce HR-EBSD maps with high PH, and $EBSP_0$ with low local MAE produce maps with low MAE. There is also considerable interaction between the local $EBSP_0$, grain's average GND density, and PH and MAE values (green boxes in Figure 4). For example, high local GND density at the $EBSP_0$ reduces the grain's average PH and increases the grain's average MAE (shown in the 'PH' and 'MAE' columns of the 'GND' rows). Therefore, using $EBSP_0$ with high MAE or low PH also leads to overestimated GND densities in the grain.

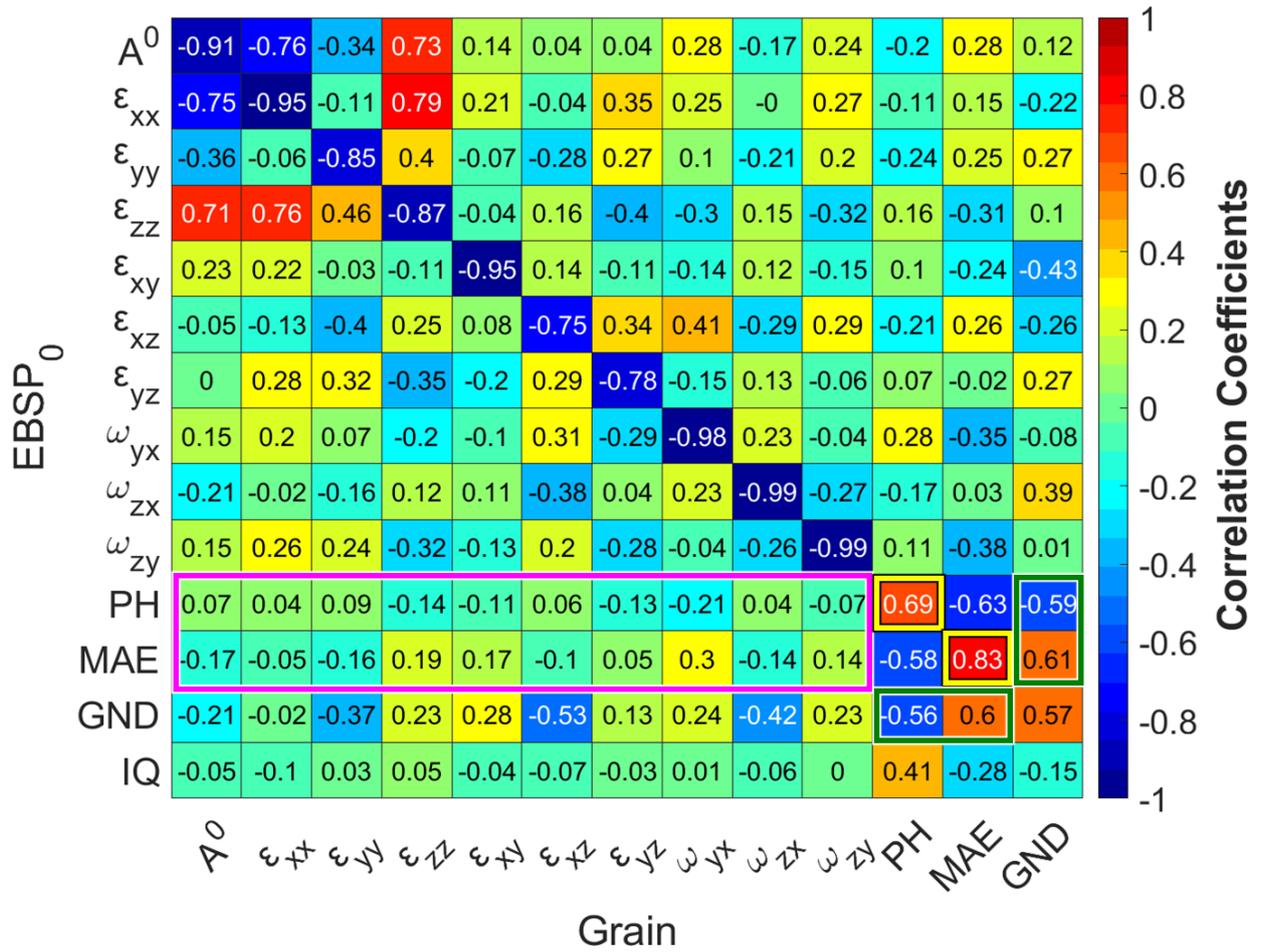

Figure 4: Correlation coefficients between the local conditions at the $EBSP_0$ point, as averaged across the other HR-EBSD maps calculated using different EBSPs (labelled $EBSP_0$), and the conditions at the grain, as averaged from all the points within the grain (labelled 'Grain'), for the ferrite grain in aged duplex stainless steel. Negative values indicate an inverse linear relationship.

All PH and MAE data points inside the grain (33 177 points) were averaged across the 51 HR-EBSD maps to infer the local response to different $EBSP_0$. These are presented in Figure 5a, and fitted by an empirical equation (3), where $a$ and $c$ are 0.014 and 0.16, respectively. The same was done for data sets from the 9 analysed grains (Figure 5b), which finds an inverse relationship between $c$ and $a$.

$$PH = \frac{a}{\sqrt{MAE}} + c \qquad\qquad 3$$

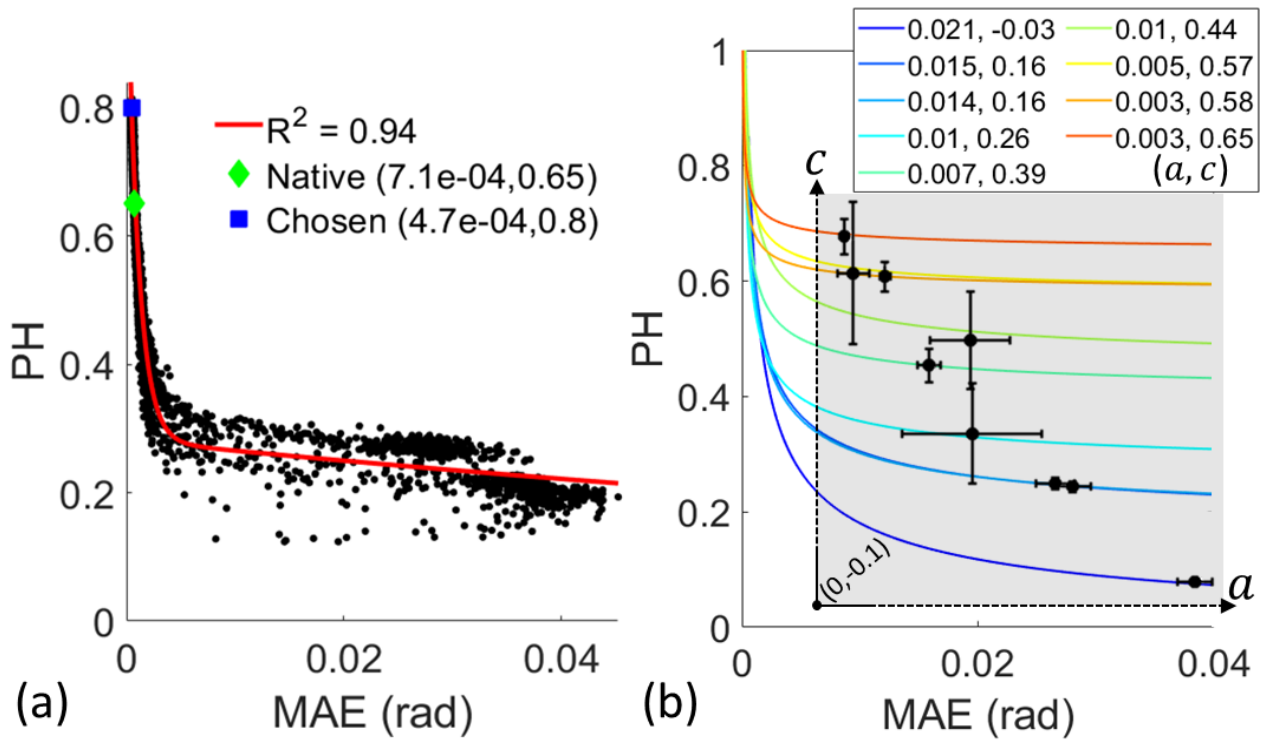

Figure 5: (a) Cross-correlation peak height (PH) and mean angular error (MAE) for all points inside the grain averaged across all different 51 HR-EBSD maps and fitted with an inverse square root function. (b) Fitting coefficients were obtained by applying the analysis to 9 HR-EBSD maps deformed in uniaxial tension and three-point bending. The variance indicates the quality of the fitting of $a$ and $c$.

Figure 4 and Figure 5 show that using the local conditions at the EBSP$_0$ as indicated by the PH and MAE can be used as a proxy metric to estimate the map's average PH and MAE values when searching for an EBSP$_0$. From this, it can be postulated that the optimum EBSP$_0$ should offer the highest cross-correlation coefficient (high PH) with the lowest fitting errors (low MAE) since such a pattern has high quality and is close to the average grain orientation. Objectively, this is the point on the EBSD map with the highest grain's average PH value and the lowest grain's average MAE value. This was identified using an iterative algorithm that finds the point with the lowest MAE in the grain and then searches for the point with the highest PH. If these are not at the same point, the 2nd lowest MAE is located and compared, then the 2nd highest PH, until these coincide at the same point. This point, termed '*Chosen*', is indicated in Figure 2, Figure 3, and Figure 5a.

The map labelled IV in Figure 1 was produced using this '*Chosen*' EBSP$_0$. The grain's average PH and MAE are 0.76 and 6 x 10$^{-4}$ rad compared to the '*Native,*' which gave a PH of 0.76 and MAE of 8 x 10$^{-4}$ rad. Although the difference in overall improvement is not significant in terms of cross-correlation parameters, the difference in strain magnitude and distribution can be seen clearly in Figure 3b, V and IV, where the difference equates to ~450 MPa (e.g., at 5 μm along the line profile).

## 3.2. Face Centre Cubic Austenite

In this example, an austenite grain in the age-hardened duplex stainless steel deformed in uniaxial tension was analysed (Figure 6). The austenite phase shows significant plastic deformation, with planar slip, compared to the harder ferrite; the mean GND density is $12.44 \pm 0.43$ $\log_{10}(\text{m}^{-2})$, compared to $11.85 \pm 0.74$ $\log_{10}(\text{m}^{-2})$ in the ferrite example (Figure 2). As for the previous ferrite example, 52 $EBSP_0$ were selected, with one being the '*Native*', 50 sampled randomly from the points with the highest IQ, and one '*Chosen*' using the algorithm to find the map point with the lowest MAE and highest PH. The locations of the points are marked on the map of GND density, which also shows significant plastic strain concentrations at some grain junctions. Due to the ductility of the austenite, larger lattice rotations occur, but the phantom strains were minimised by the re-mapping process [24]. The examples of strain maps (in-plane shear component) show the dependence of the strain magnitude and sign on the choice of reference pattern (Figure 6 and Figure 7, in which the arrow marks a subgrain boundary with low misorientation < 0.5°).

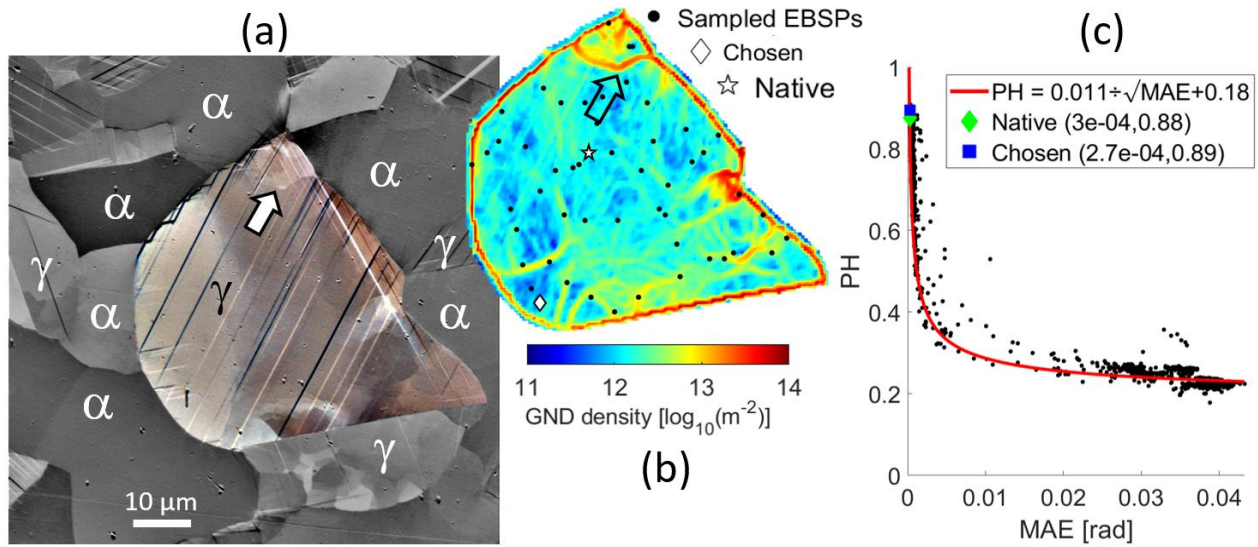

Figure 6: (a) Forescatter electron (FSD) image with the other grains grey-scaled to accentuate the grain being examined. (b) Estimated geometrically necessary dislocations (GND) density map with the location of the 50 randomly selected $EBSP_0$ plus '*Native*' and '*Chosen*'. (c) Cross-correlation peak height (PH) and mean angular error (MAE) for all points inside the grain were averaged across all 51 HR-EBSD maps and fitted with an inverse square root function to find the '*Chosen*' pattern. The goodness of fit ($R^2$) equals 0.79. The sample is at 4% engineering strain; see the supplementary data: B.

As in the analysis of the ferrite grain, the overall distribution of the in-plane shear strain is similar regardless of the reference choice, as highlighted by the strain peak at the sub-boundary (highlighted by an arrow in Figure 6 and Figure 7). However, the finer details of the distribution of strain and magnitude depend on the $EBSP_0$; for instance, a deformed point near the ferrite-austenite grain boundary (4th point in Figure 7a.IV) renders the grain field as being negative even when the applied deformation was in tension, and 'Native' (IV line) behave very different from 'Chosen' (line VI) at 28 µm and 35 µm. A similar observation can be made for the 1st point (Figure 7a.I), but the field is less negative, and a more positive in-plane shear is induced when the point is selected inside the sub-grain (3rd point). In general, the difference in distribution, as in the previous example, is not due to an offset but an observable change. The GND density is not very sensitive to the reference point (Figure 7b); however, distribution fluctuated, especially at 32 µm.

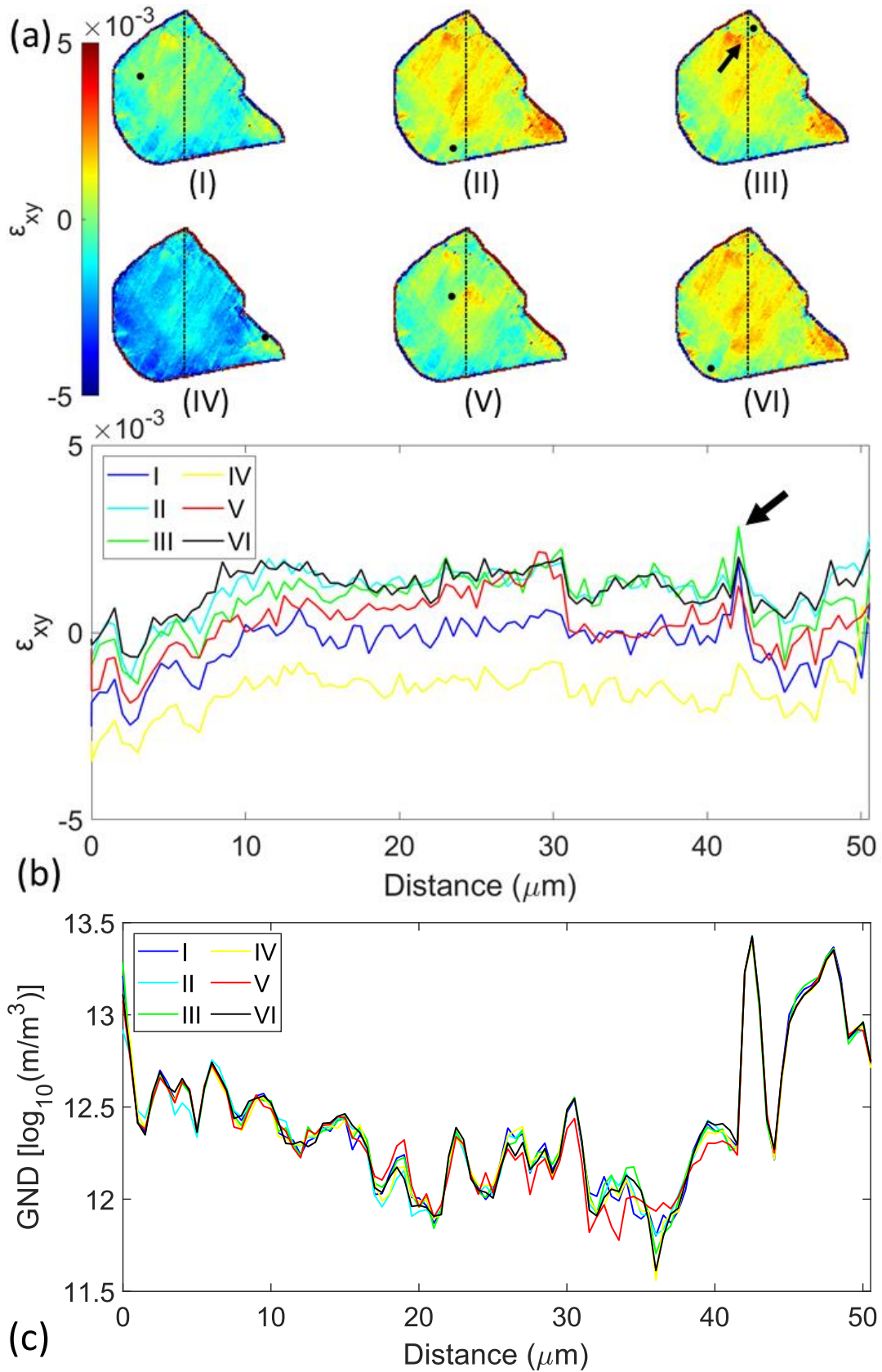

Figure 7: a) In-plane shear strain ($\varepsilon_{xy}$) fields produced using six different $EBSP_0$ in an austenite grain. A black dot indicates the position of the reference pattern. The line profile along the dotted for (b) the shear strain and (c) GND density. The arrow points to the sub-boundary, also seen in Figure 6. V and VI are '*Native*' and '*Chosen*' $EBSP_0$. The IQ values of points I to VI are 0.4225, 0.4619, 0.4569, 0.4956, 0.4737 and 0.4794.

The same correlative analysis used for the ferrite (Figure 4) was applied to 3 different grain maps of the face centre cubic austenite (204 maps at different engineering strains – see the supplementary information: B). The correlation coefficients between the local conditions at the $EBSP_0$ and the grain are presented in Figure 8 by averaging the correlation coefficients from 5 sets of analysis. This found weaker diagonal relationships between the in-plane shear strain (and consequently the stress) tensor (as indicated by the -0.32 correlation coefficient). The inverse correlation between the grain's and the reference pattern's strain components along the diagonal is less pronounced than in the ferrite phase. The local MAE strongly influences the resultant grain's averaged MAE, PH, and estimated GND density. Still, the influence of IQ of the reference pattern on the resultant field, similarly to the ferrite example, is lower.

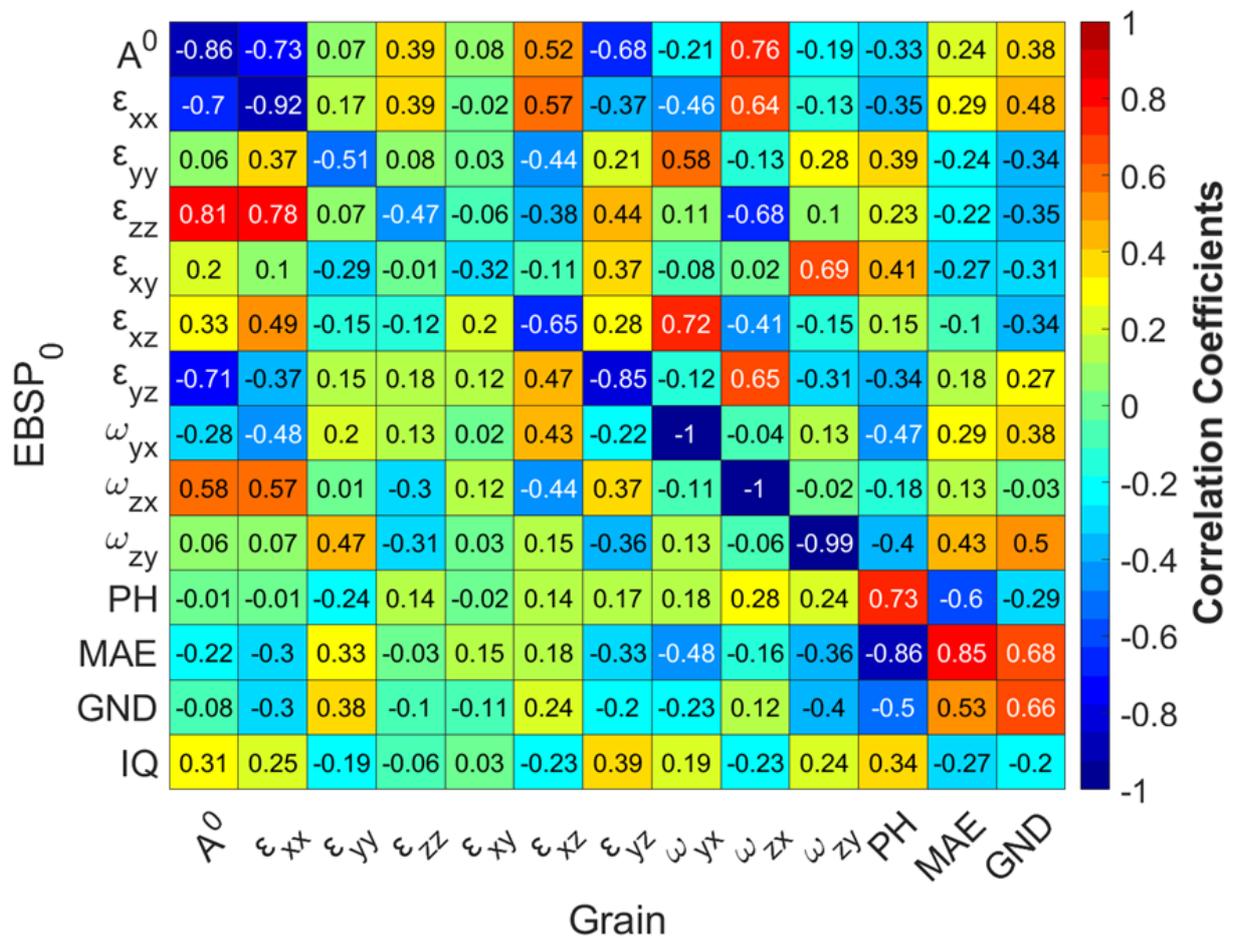

Figure 8: Correlation coefficients between the local conditions at the $EBSP_0$ point, as averaged across the other HR-EBSD maps calculated using different EBSPs (labelled $EBSP_0$), and the conditions at the grain, as averaged from all the points within the grain (labelled 'Grain'), for the austenite grain in aged duplex stainless steel. Negative values indicate an inverse linear relationship.

Fitting equation 3 to the PH and MAE data for the austenite grain gave $a$ and $c$ of 0.011 and 0.18, respectively (Figure 6c). The 'Chosen' $EBSP_0$ had an average local PH of 0.89 and MAE of $2.7 \times 10^{-4}$ rad compared to the '*Native*' local PH of 0.88 and MAE of 3 x 10⁻⁴. The grain's average PH and MAE for the '*Chosen*' $EBSP_0$ were 0.85 and 13 x 10⁻⁴ rad, marginally different from the grain's average values using '*Native*,' which gave a PH of 0.85 and MAE of 14 x 10⁻⁴ rad. However, as for the ferrite example, although the mean improvement in correlation parameters was not substantial, the effect on the stress magnitudes and distribution cannot be ignored, which is observed in Figure 7, comparing V and VI, where the difference in the in-plane shear strain equates to ~250 MPa.

### 3.3. Silicon

EBSD maps were collected for the silicon sample around the tip of a cleavage crack as it was propagated in a quasi-static manner in 12 intervals [89]. The change between the stress maps due to different $EBSP_0$ was minimal in magnitude and distribution once the $EBSP_0$ was selected to be remote (> 12 μm) from the crack (Figure 9a). Applying the same correlative analysis to the 12 observations (12 datasets x 52 sampled $EBSP_0$ = 624 maps) gives Figure 10. The diagonal correlation for the <span style="color:red">lattice distortion</span> components is similar to that observed in the ferrite, with an inverse correlation between the grain and $EBSP_0$ local average value in each tensor of the strain and rotation components. This strong correlation is also seen in MAE, but EBPS0 image quality, PH, MAE, or the choice of $EBSP_0$ do not influence the estimated GND density, because, at room temperature, silicon has no significant plasticity [90], <span style="color:red">and the dislocation density is lower than the HR-EBSD GND density measurement noise floor</span> [28].

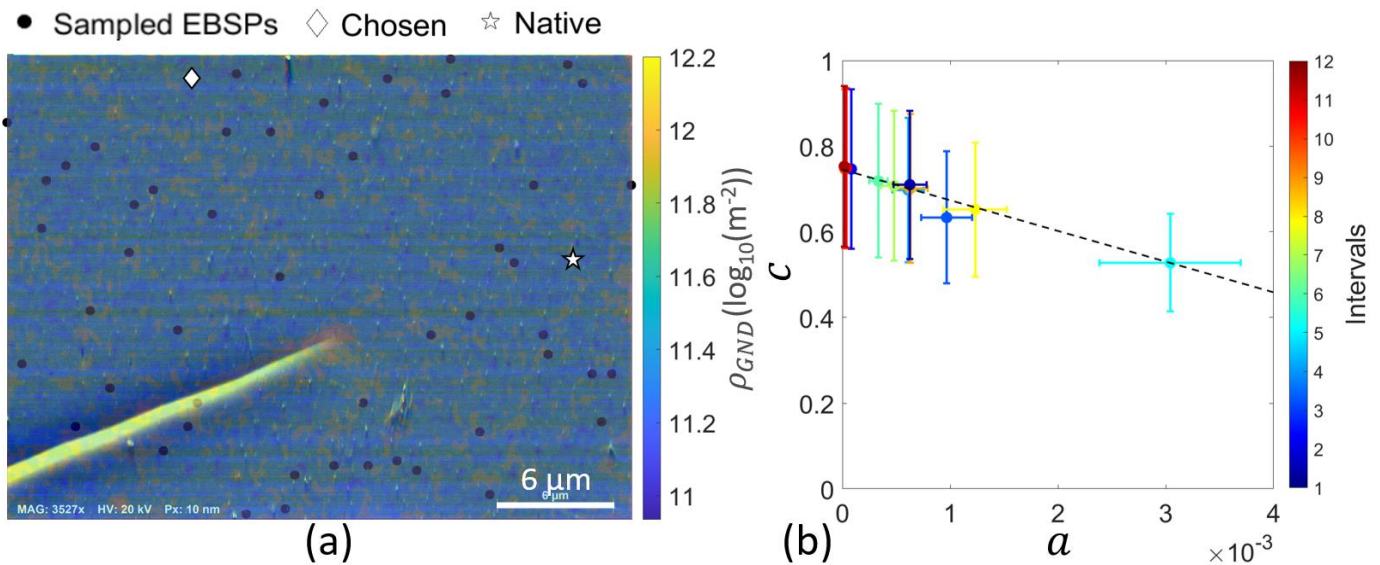

Figure 9: (a) Forescatter electron (FSD) image for the crack number propagating in (001) silicon-crystal deformed in compression overlayed with the estimated GND density map that includes the location of the 52 $EBSP_0$ including '*Native*' and '*Chosen*'. (b) Fitting coefficients $a$ and $c$ from equation (4) with the legend indicating the experiment intervals.

The '*Chosen*' EBSP$_0$ was taken as the point with the highest PH and lowest MAE, and a trend was observed between the fitting parameters for the 12 datasets as in the previous analysis of the ferrite (Figure 10). Using the '*Chosen*' compared to '*Native*' did not significantly reduce the elastic strain but slightly improved PH and MAE in the calculated field (see the supplementary information: C).

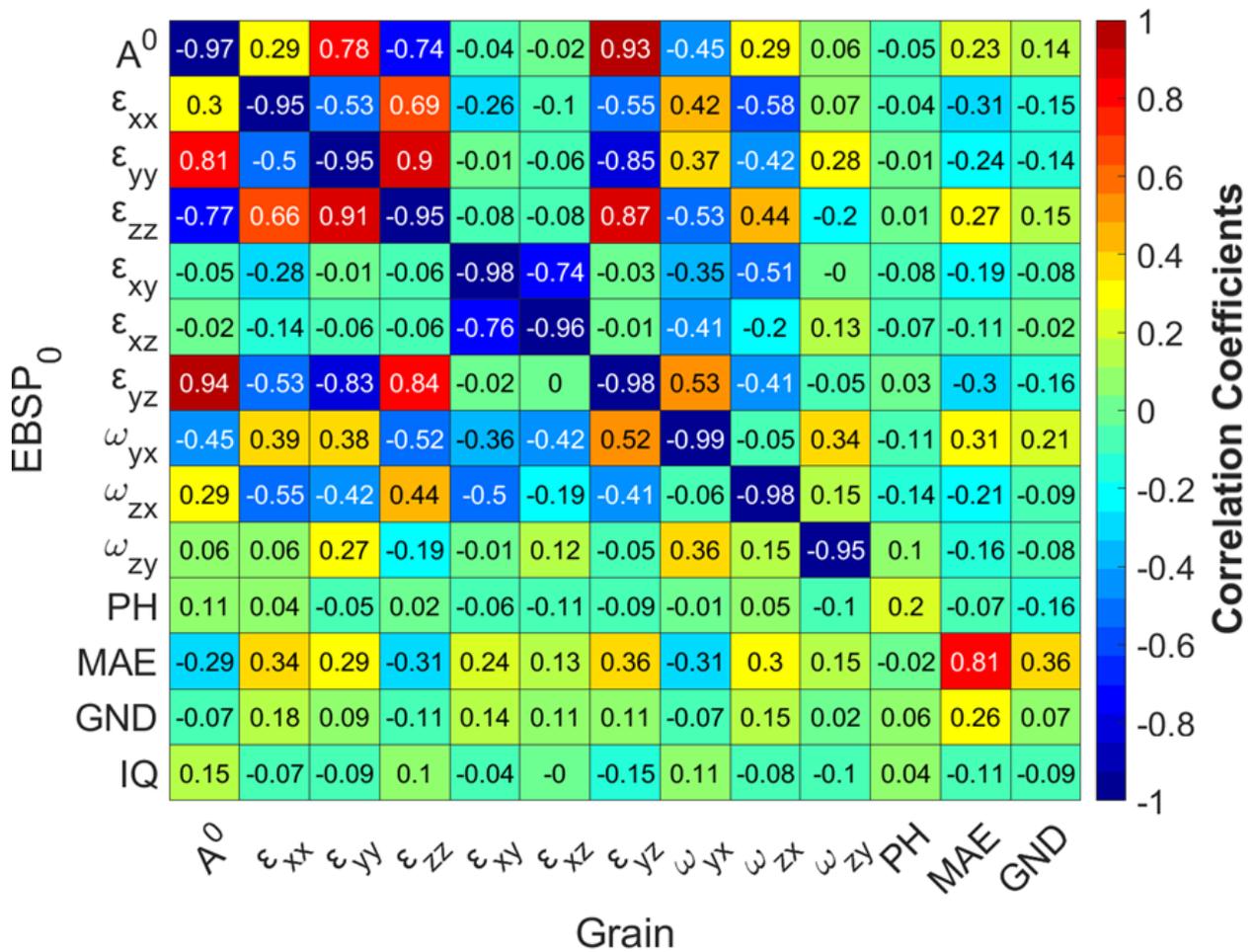

Figure 10: Correlation coefficients for the grain average (Grain) and reference point average (EBSP$_0$) averaged for 12 Si maps. The minus sign indicates an inverse relationship.

## 4. Discussion

### 4.1. The effect of EBSP₀

The correlation coefficients are summarised in Figure 11 for the elastic silicon (Si) sample and the ferrite (Fe-α) and the austenite (Fe-γ) phases, which deformed plastically to different degrees. The figure shows correlation coefficients between the local conditions at the EBSP₀ point and the averaged conditions at the grain for different parameters: the average elastic deformation gradient tensor ($A^0$) determinant, maximum in-plane principal strain ($\varepsilon_{Max}$), rotation magnitude ($\omega_T = \sqrt{\omega_{32}^2 + \omega_{13}^2 + \omega_{21}^2}$), peak height (PH), mean angular error (MAE) and GND density, and materials: the ferrite (Fe-α) and austenite (Fe-γ) phase of age-hardened DSS, and Silicon (Si). Figure 11 shows that correlations between the local conditions at the EBSP₀ point and the averaged conditions at the grain are material-dependent and impact the calculated field parameters.

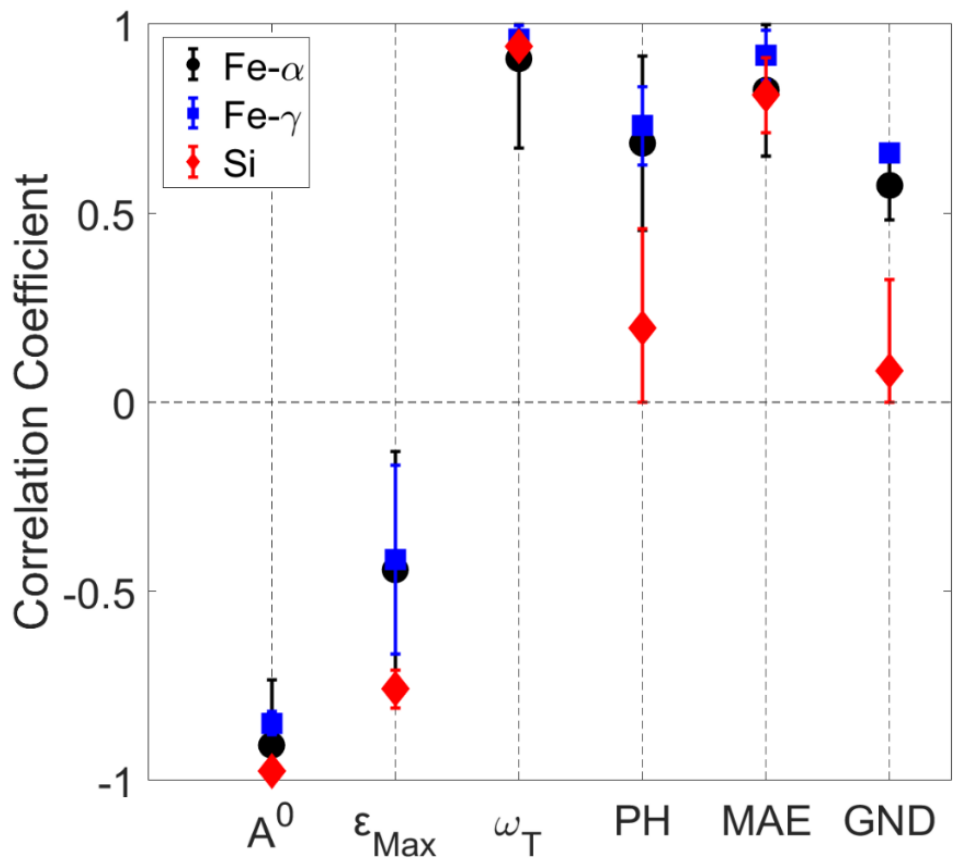

Figure 11: Linear correlation coefficients between the local conditions at the EBSP₀ point and the averaged conditions at the grain for the ferrite (Fe-α) and austenite (Fe-γ) phase of age-hardened DSS, and Silicon (Si). The analysis considers the average deformation gradient tensor determinant ($A^0$), maximum in-plane principal strain ($\varepsilon_{Max}$), rotation magnitude ($\omega_T = \sqrt{\omega_{32}^2 + \omega_{13}^2 + \omega_{21}^2}$), correlation peak height (PH), mean angular error (MAE) and GND density.

The full-field elastic deformation gradient and the strain strongly depend on the absolute lattice distortion at the EBSP$_0$, which is expected for a relative strain measurement. However, this correlation weakens with increasing plastic deformation in the grain, which shows that the EBSP$_0$ selection is still important even if only relative strains are considered, especially if the material is plastically deformed. However, the rotation magnitude ($\omega_T$) and MAE have a high positive correlation coefficient unaffected by the material plasticity. This confirms that minimising the local EBSP$_0$'s MAE is always a good way to minimise the grain's MAE in both plastically and elastically deformed materials.

In the plastically deformed grains, the grain's average GND density is positively correlated with local EBSP$_0$ GND density, even though GND density is nominally considered insensitive to EBSP$_0$ selection. Within the set of randomly selected reference points with high IQ, the EBSP$_0$ seems to affect the GND density distribution and magnitude, which contradicts the assumption that HR-EBSD can precisely determine the lattice distortion gradient independently of the EBSP$_0$ [40,61,62]. EBSP$_0$ selection can also cause individual features in the GND density map to appear or disappear (compare line profiles of Figure 7c). However, the GND density is typically reported in log scale maps (Figure 7c); thus, this influence (roughly less than $\pm$ 0.04 log$_{10}$ (m/m$^{-3}$)) is relatively small.

The degree of correlation between local-EBSP$_0$'s MAE and PH with the grain's average GND density is material-dependent. For example, the ferrite showed a strong correlation between the resultant map's PH and MAE with the EBSP$_0$ image quality (IQ) and GND density, where an EBSP$_0$ with high IQ and low GND density produced a map with high PH and low MAE. This is consistent with the widespread use of the '*Native*' selection (which selects EBSP$_0$ with high IQ and low GND density) across the literature [30,62,91–95], as this is likely to produce EBSD maps with relatively high PH and low MAE. However, this correlation is weaker for the austenite grain and is negligible for the Si crystal.

Thus, the lattice distortion arising from residual elastic strains and crystal defects within the electron interaction volume at the reference point affects the PH and MAE, as crystal defects such as dislocations reduce PH and increase MAE. In contrast, residual elastic strain mainly increases MAE by changing the position of zone axes within the diffraction pattern. This further shows the significance of local lattice rotation-induced perturbations on patterns, which increase with GND density. Note that the correlation coefficient discussed here (i.e., Pearson correlation coefficient) is a linear correlation coefficient that ignores non-linear components of a correlation relationship that may exist.

The choice of $EBSP_0$ changes the spatial distribution of both lattice distortion and GND density maps. This is contrary to the adopted assumption that all relative lattice distortion values are only shifted from the absolute distortions by a constant equal to the absolute distortion at $EBSP_0$ (which could be identified, for example, using a CPFE simulation [56,57]. Recently, an exciting approach to studying this phenomenon was made by simulating an EBSP from a deformed volume of interaction [18,96]. The reverse engineering of the problem by applying a lattice distortion to the EBSP helps separate what each deformation component does to the EBSP geometry, the deformation's effect on intensity distributions in the EBSP, pattern distortion from instruments, and can provide insights into how the correlation between sharp and blurred patterns can be used to estimate plastic strains.

Furthermore, there is no apparent connection between $EBSP_0$'s IQ and $EBSP_0$'s local lattice distortion. This is contrary to qualitative studies that used pattern characteristics as indicators of elastic strain gradient [97]; however, we only considered high-quality patterns, and Wright *et al.* [98] asserted that due to the relationship between electron beam spot size and probed pattern degradation, qualitative analysis that uses IQ is only adequate for a tungsten filament scanning electron microscope and not a FEG (field emission gun) microscope, which was used here.

When the PH and MAE of the grain were averaged across the 51 HR-EBSD maps, the result could be fitted to an empirical equation (eq. 3) where the PH is related to the inverse of the square root of MAE through two coefficients, $a$ and $c$. This inverse relationship between the PH and MAE encapsulates the effect of the stored dislocations on the blurring of the patterns and the impact of lattice distortion on the shifts in the zone axes. The optimum $EBSP_0$, i.e., the '*Chosen*' reference points, should offer the lowest errors (low MAE) with the highest cross-correlation (high PH). Such a pattern naturally has high quality and is close to the average grain orientation. High PH indicates low crystal defect density within the interaction volume, which blurs the EBSD pattern due to local lattice distortions near defect centres such as dislocation cores. In the case of GNDs, a lattice rotation gradient is also produced across the interaction volume [99]. Low MAE relates to the consistent description of the strain state, i.e., it reduces the uncertainty that arises from the unknown absolute lattice distortion at the $EBSP_0$ on the calculated lattice distortion field by improving the HR-EBSD measurement precision (less random point-to-point noise). Thus, the search algorithm applied to select the optimum $EBSP_0$ gives more weight to MAE as the induced noise (from pattern acquisition or stored crystal defects) is directly affected by the local MAE. PH is also more affected by MAE as plasticity increases, with the correlation between local MAE and grain PH increasing from -0.13, -0.58, and to as high as -0.86 for the Si, Fe-$\alpha$, and Fe-$\gamma$, respectively. The need for re-mapping increases due to significant lattice rotation; thus, the suitability of the $EBSP_0$ becomes more critical while reducing phantom strains [38,39].

Finally, the offset-corrected standard deviation between the six $\varepsilon_{xy}$ line profiles in Figure 1, without considering the measurement's mean angular error (MAE) which is a cross-correlation parameter that quantifies the imprecision in fitting the HR-EBSD pattern's shifts to a deformation gradient tensor, is ± 7.3 × $10^{-4}$ (Supplementary Figure 2). Since it is within the same magnitude of the typical HR-EBSD measurement mean (angular) error (MAE), this might imply that the spatial distribution's offset-corrected variance can be described using the MAE to quantify the noise induced due to $EBSP_0$ selection. But to make this assumption, we will need to (1) calculate the spatial distribution's offset-corrected standard deviation between HR-EBSD maps created using different $EBSP_0$ and an accurate (absolute) strain map, (2) calculate the MAE of each HR-EBSD map created using different $EBSP_0$. The correlation between (1) and (2) – for a statistically significant number of maps – will indicate whether the HR-EBSD's MAE can sufficiently describe the spatial distributions' variance; thus, the precision. However, verifying whether the MAE of different $EBSP_0$-based maps is a true measure of HR-EBSD precision, compared to a true measurement, has never been done, and can bring much-needed clarity about HR-EBSD (true) precision.

## 4.2. Selecting an $EBSP_0$

We defined the optimal reference pattern as the one that enables the most precise measurement of the deformation gradient tensor, i.e., produces the lowest average MAE in the grain [81]. Although the data being analysed was measured in situ at different elastic and plastic straining levels, the method explored here is indifferent to whether the material is relaxed or loaded. Therefore, the method presented in this work is to select a reference pattern to maximise the precision of the elastic strain measurements, not the accuracy, which depends on the absolute strain state of the reference pattern, and that is generally unknown.

This approach operates under the substantiated hypothesis[4] that a good $EBSP_0$, which produces an HR-EBSD map with low grain's average MAE and high grain's average PH values, tends to show up as a point with low local MAE and high local PH in the HR-EBSD map computed with respect to a different $EBSP_0$. The $EBSP_0$ selected using the MAE and PH-based criteria were termed '*Chosen*' in this work. In practice, the $EBSP_0$ search algorithm uses both the MAE and PH in the optimisation process; the strong inverse correlation between these two parameters (as shown in Figure 5, Figure 6) means there is no harm in using them both, and this could improve the robustness of the selection process by excluding rogue patterns with poor cross-correlation quality, but erroneously precise solutions (e.g., measuring cross-correlation shifts of all zeros).

The optimal $EBSP_0$ offers (1) maximum EBSD cross-correlation quality; and (2) the most precise fit to the deformation gradient tensors from the measured ROI shifts across the entire grain, i.e., all EBSPs which use it as a reference point. Expressed as HR-EBSD quality metrics, this is the $EBSP_0$ with (1) the highest grain's average PH; and (2) the lowest grain's average MAE. Using these selection criteria, a brute-force search for the optimal $EBSP_0$ would require the computation of as many HR-EBSD maps as there are points in the grain, which would be computationally unfeasible. Even interrogating 51 random $EBSP_0$, as we did in this paper, is very computationally expensive.

A more viable approach uses an iterative algorithm starting with the '*Native*' $EBSP_0$. The $EBSP_0$ for each iteration is selected from the local MAE and PH values averaged over all previous map iterations until the grain's average MAE and PH values converge at a low minimum value. This approach is more computationally efficient. As shown in Figure 12, the optimality of the $EBSP_0$ is substantially increased just one iteration after the initial '*Native*' $EBSP_0$, with a suitable $EBSP_0$ being found after only three iterations with a resultant map that is comparable to one produced from an $EBSP_0$ found after 51 iterations, i.e., '*Chose*n' $EBSP_0$.

---

[4] In other words, there is a strong correlation between the grain's average MAE and PH (of a selected $EBSP_0$), and its local MAE and PH (computed using any other $EBSP_0$). The strength of these correlations is shown for three different crystal types in the Results section, in Figure 4, Figure 8, and Figure 10.

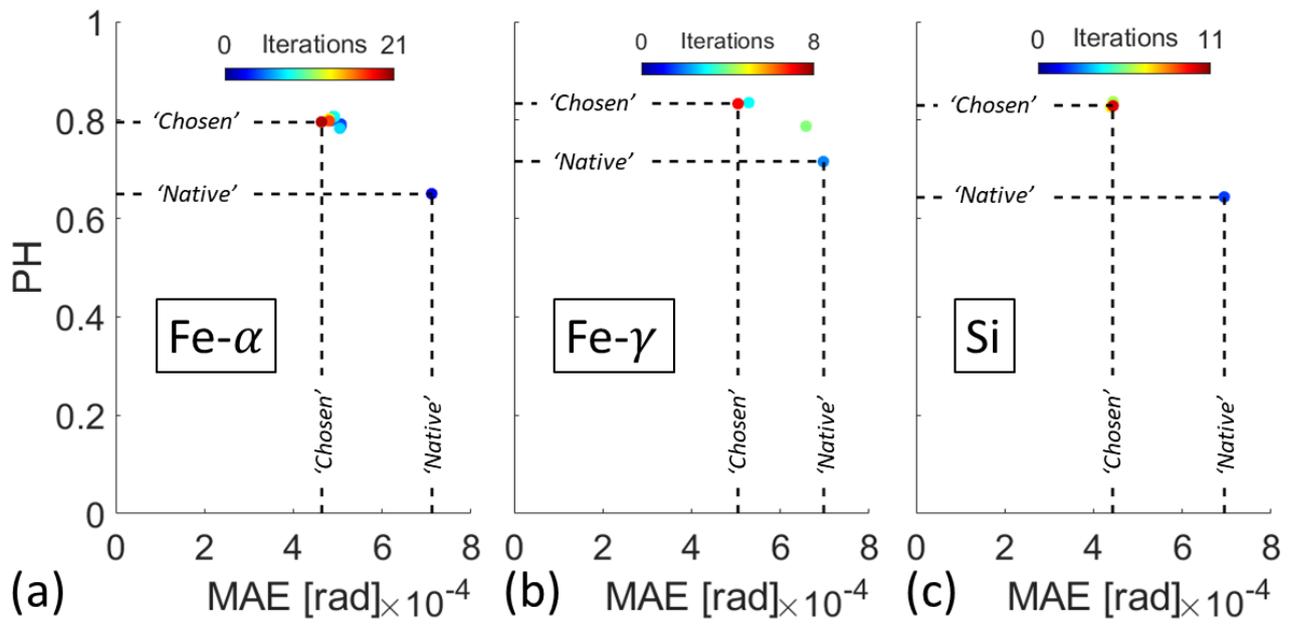

Figure 12: Number of iterations required to find the 'Chosen' $EBSP_0$ starting from a 'Native' $EBSP_0$ for (a) ferrite, (b) austenite, and (c) silicon.

## 5. Conclusion

The effect of the reference pattern (EBSP$_0$) on HR-EBSD analysis of deformation in cubic crystals with different degrees of plastic deformation has been studied to conclude that the local lattice distortion at the EBSP$_0$ influences the resultant HR-EBSD map, e.g., a reference pattern deformed in tension will directly reduce the HR-EBSD map tensile strain magnitude while indirectly influencing the other component magnitude and the strain's spatial distribution. Furthermore, the choice of EBSP$_0$ slightly affects the GND density distribution and magnitude, and choosing a reference pattern with a higher GND density reduces the cross-correlation quality, changes the spatial distribution and induces more errors than choosing a reference pattern with high lattice distortion. Additionally, there is no apparent connection between EBSP$_0$'s IQ and EBSP$_0$'s local lattice distortion.

The study revealed an empirical relationship between the cross-correlation peak height (PH) and the mean angular error (MAE), in the form of $\mathrm{PH} = a/\sqrt{\mathrm{MAE}} + c$, where $a$ and $c$ are fitting parameters. This relationship was used – iteratively – to find the optimal EBSP$_0$, which improved the HR-EBSD precision. Although our method does not enable absolute strain measurement in HR-EBSD; it does increase the precision of lattice distortion maps and might complement methods based on enhancing the probed field by independently approximating the strain at the EBSP$_0$ point.

## Acknowledgements

We thank Dr Roger Francis (Rolled Alloys www.rolledalloys.com) and Professor Peter Wilshaw (University of Oxford) for supplying the specimen materials. The authors acknowledge the use of characterisation facilities within the David Cockayne Centre for Electron Microscopy (DCCEM), Department of Materials, University of Oxford, alongside financial support provided by the Henry Royce Institute (Grant ref EP/R010145/1). Abdalrhaman Koko is supported by an EPSRC research studentship (Grant ref EP/N509711/1).

## Authorship Contribution Statement

**Abdalrhaman Koko:** Conceptualization, Methodology, Software, Investigation, Formal analysis, Writing - original draft, Visualization.

**Vivian Tong:** Writing - review & editing.

**Angus J. Wilkinson**: Software, Writing - review & editing, Supervision.

**Thomas James Marrow:** Resources, Writing - review & editing, Supervision, Funding Acquisition.